# Axial range of conjugate adaptive optics in two-photon microscopy

Hari P. Paudel,[1,*] John Taranto,[2] Jerome Mertz,[3] and Thomas Bifano[4]

[1]*Department of Electrical and Computer Engineering, Boston University, 8 Saint Marys St., Boston, MA 02215, USA*
[2]*Thorlabs Inc. 56 Sparta Ave, Newton, NJ 07860 USA*
[3]*Department of Biomedical Engineering, Boston University, 44 Cummington Mall, Boston, MA 02215, USA*
[4]*Department of Mechanical Engineering, Boston University, 110 Cummington Mall, Boston, MA 02215, USA*
*paudelhp@bu.edu*

**Abstract**: We describe an adaptive optics technique for two-photon microscopy in which the deformable mirror used for aberration compensation is positioned in a plane conjugate to the plane of the aberration. We demonstrate in a proof-of-principle experiment that this technique yields a large field of view advantage in comparison to standard pupil-conjugate adaptive optics. Further, we show that the extended field of view in conjugate AO is maintained over a relatively large axial translation of the deformable mirror with respect to the conjugate plane. We conclude with a discussion of limitations and prospects for the conjugate AO technique in two-photon biological microscopy.

## 1. Introduction

Multiphoton microscopy has become an important technique for imaging deep within biological tissue because of its selectivity to ballistic excitation photons in comparison to those that are scattered [1]. Nevertheless, aberrations at the tissue interface or within the tissue itself lead to reduced confinement of the focused excitation spot. This in turn diminishes signal intensity and limits achievable imaging depth. This problem of aberration-induced signal loss is more pronounced in higher-order multiphoton microscopy, which otherwise has the potential for much deeper imaging [2–5].

Adaptive optics (AO) is one approach to compensating these aberrations in microscopy [6–11]. The idea of AO is to introduce a wavefront control element, such as a deformable mirror (DM), to compensate wavefront distortions generated by sample-induced aberrations. In a scanning microscope, such as a two-photon microscope, this control element is inserted in the excitation beam path, most commonly in a plane conjugate to the back aperture, or pupil, of the objective [12–21]. We refer to that configuration as pupil AO.

In principle, pupil AO is effective at correcting spatially (or shift) invariant aberrations in the system; however, as is well known from astronomical imaging [22–25], it is less effective at correcting spatially variant aberrations, in which case it leads to restricted fields of view (FOV). To correct for spatially variant aberrations, a more effective placement of the DM is in a plane conjugate to the primary source of aberrations, called conjugate AO (generalized to multi-conjugate AO in the case of multiple aberration planes and corresponding conjugate DM planes [26–31]). The FOV advantage of conjugate AO in microscopy applications has been studied using numerical simulations [27,29,30]. It has also been demonstrated experimentally in linear microscopy applications, both scanning [26,28] and widefield [31]. We report here a demonstration of conjugate AO in a nonlinear (here two-photon) microscopy application. Our demonstration is restricted to the simplified geometry of 2D sample and well-defined interface aberrations located at a plane of known separation from the sample. As such, it is a proof of principle demonstration intended to explore some limitations of conjugate AO. Specifically, we examine the axial range of conjugate AO correction, as a step toward generalization of its application to volumetric samples with axially distributed aberrations.

## 2. Experimental method

A schematic of our two-photon microscope, capable of both pupil and conjugate AO with two independent DMs, is illustrated in Fig. 1.

The excitation source is a 2.9 Watt, 140 fs, 80 MHz repetition rate Ti-Sapphire laser (Coherent Chameleon), operated at 880 nm. The laser power is controlled by a motorized half-wave plate (Thorlabs AHWP05M-980) and polarization beam splitter (Thorlabs GT5-B). Two pairs of doublet achromatic lenses, f1=145 mm and f2=245 mm, conjugate two orthogonally scanning galvanometric mirrors (Thorlabs GVS011) to the pupil DM (PDM: Boston Micromachines Corp. Kilo-DM, 1020 segmented actuators, >10 kHz update rate, 1.5 $\mu$m stroke), itself conjugated to the back aperture (pupil) of the microscope objective (Nikon N16XLWD-PF 16×, NA=0.8, WD=3mm).

In addition to providing pupil AO, our system can provide conjugate AO, which can be engaged with the help of two flip mirrors (FM). When engaged, additional relay optics are introduced in the excitation optics of the microscope, comprising a pair of doublet achromatic lenses (f2=245 mm), a central polarization beam splitter (PBS: Thorlabs PBS252) located in a pupil plane, two quarter-wave plates (Thorlabs WPQ10M-850) and two biconvex lenses (f3=40 mm). Also included in this relay is the conjugate DM (CDM: Boston Micromachines Corporation KiloDM, 1020 actuators, > 20 kHz update rate, 3 $\mu$m stroke), mounted on a translatable carriage, along with a compensation mirror (CM), such that the distance between the CDM and CM is maintained fixed at 160 mm. The purpose of the translatable carriage is to allow the position of the CDM to be adjusted so that it can be conjugated to a range of axial positions between the microscope focal plane (where the sample is located) and the front window of the microscope objective. The purpose of the compensating mirror is to maintain a fixed path length throughout the relay optics such that the introduction of the conjugate AO produces a net unit magnification independent of the position of the translatable carriage.

Upon operation of the microscope, two-photon fluorescence produced by the sample is collected in an epifluorescence mode and routed with a dichroic mirror (Semrock FF665-Dio2), collection lens, and emission filter (Thorlabs MF525-39) to a photomultiplier tube (Hamamatsu H7422), whereupon the photocurrent is amplified by a transimpedance preamplifier (Thorlabs TIA60) and digitized by a 14 bit digitizer (Alazar ATS460 125 MS/s). The digitizer is operated in an external trigger mode for fast data transfer synchronized to the update clock of the DM driver (PDM or CDM), or to a frame clock generated by a DAC card (NI PCIe 6232).

To perform a proof-of-principle demonstration of our AO system, we purposefully introduced aberrations in our system in the form of a phase screen. This phase screen was produced using a grayscale laser mask writer (Heidelberg DWL66). Specifically, a 2D sinusoidal pattern of peak-to-valley height 3 $\mu$m and period 200 $\mu$m was created by rastered laser exposure of a 30 $\mu$m thick layer of AZ P4620 photoresist coated onto a 300 $\mu$m thick glass substrate. After fabrication, three dimensional geometry of the phase screen was measured using the Zygo NT6000 white-light interferometer. An additional 100 $\mu$m thick microscope coverslip was placed on top of the patterned photoresist to protect it during use.

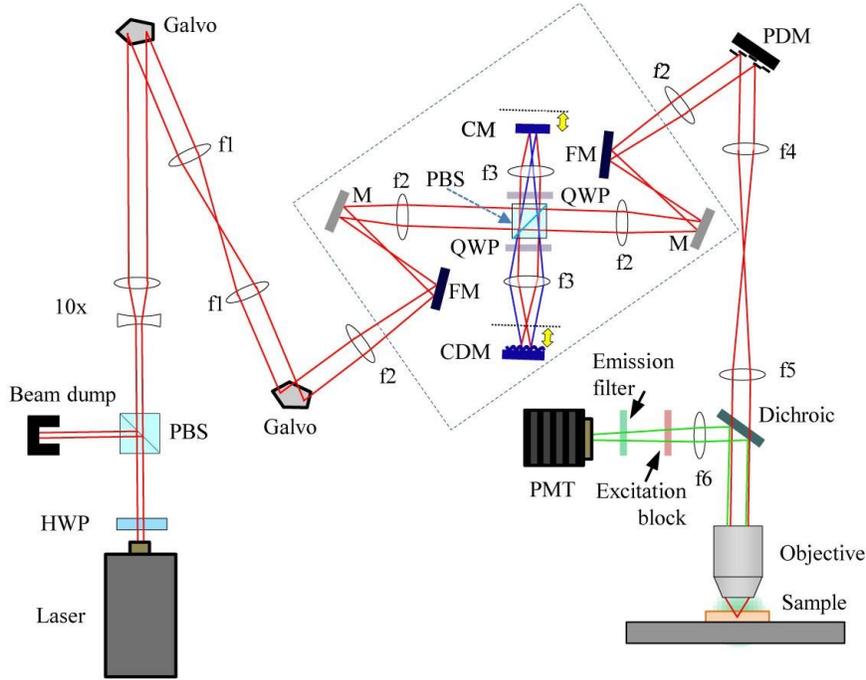

Fig. 1. Schematic of a two-photon microscope with pupil AO and conjugate AO. HWP=half wave plate, QWP=quarter wave plate, PBS=polarizing beam splitter, M=mirror, FM=flip mirror, and PMT=photomultiplier tube, PDM=pupil deformable mirror, CM=conjugate mirror, CDM=conjugate deformable mirror, and f1-f6=lenses. Optics enclosed in the dashed box comprise the conjugate AO component of microscope. Two thick arrows indicate the displacement of CDM and CM from image planes (indicated by dashed lines) to the aberration conjugate planes. Components in blue indicate parts mounted on a common motorized translation stage. Rays in blue illustrate representative changes depending on the position of the conjugate plane.

To compensate for the aberrations introduced by our phase screen, we used image-based iterative feedback optimization, where the fluorescence intensity served as the optimization metric. For pupil AO correction, we parked the excitation focus at the center of the sample and used a sequential optimization technique with 1024 Walsh orthogonal modes, the details of which are described in [32]. For conjugate AO, we scanned the beam over the entire image FOV and optimized the total fluorescence intensity per image based on a stochastic parallel gradient descent (SPGD) algorithm [33]. While conjugate AO optimization could have been performed by acquiring full raster-scanned images at each iteration step, we found we could significantly increase the speed of our optimization (by two orders of magnitude) by instead acquiring sparse representations of these images using a much faster Lissajous scan pattern.

Finally, we note that attempts to perform pupil AO using fluorescence acquired from full images rather than from a single point did not lead to any fluorescence increase or image enhancement, as expected from the fact that pupil AO provides spatially-variant aberration correction over only limited FOVs [31].

## 3. Results

To test the capacity of conjugate AO to perform aberration corrections over a large FOV, we imaged a sample consisting of a single layer of 1 $\mu$m fluorescent beads (Fluoresbrite, Polysciences) attached to a microscope slide. The separation distance between the fluorescent beads and the aberrating phase screen was $d=300$ $\mu$m. To properly conjugate the CDM to the phase screen, we displaced it from the nearest intermediate image plane (see Fig. 1) by a distance $M^2d$, where $M$ is the (telecentric) magnification from the phase screen to the CDM (here 6.4×), leading to a CDM translation distance of 12 mm from the intermediate image plane. Conjugation of the CDM to the phase screen was independently verified by inserting a camera in a conjugate plane (not shown in Fig. 1). Vignetting caused by the fold mirrors in our confined optical setup limited the maximum FOV of our microscope to about 250 $\mu$m × 250 $\mu$m. An aberrated image of fluorescent beads is shown in Fig. 2(a), where, manifestly, the aberrations due to the phase screen caused the beads to be unresolvable. Images taken after conjugate and pupil AO correction are shown in Figs. 2(b) and 2(c), respectively (in both cases, the non-active DM was set to a flat state). Higher resolution images (100 $\mu$m × 100 $\mu$m) are shown in Figs. 2(d)-2(f). As is apparent, conjugate AO correction is effective over the entire (albeit vignetted) FOV of our microscope, whereas pupil AO is effective over only a narrow FOV about the image center.

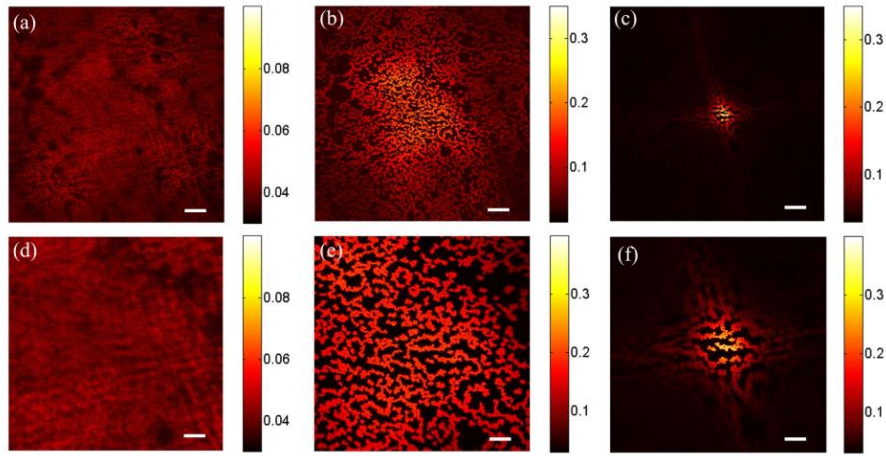

Fig. 2. Fluorescent beads (1$\mu$m diameter) in a 250 $\mu$m × 250 $\mu$m FOV imaged through the phase screen, (a) without correction, (b) with conjugate AO correction, and (c) with pupil AO correction. Higher resolution images (100 $\mu$m × 100 $\mu$m FOV) are also shown (d) without correction, (e) with conjugate AO correction, and (f) with pupil AO correction. Scale bars are 25 $\mu$m for images (a)-(c) and 10 $\mu$m for images (d)-(f)

In Fig. 3, we compare the measured aberration topography map of the phase screen (Fig. 3(a)) with the final shapes applied to the conjugate (Fig. 3(b)) and pupil (Fig. 3(c)) DMs. We recall that the wavefront map amplitude $W(x)$ is twice the topography map in reflection mode and $(n-1)$ times the topography map in transmission mode, where $n$ is the index of refraction of the aberration substrate; the corresponding phase map is related to wavefront map by $\varphi(x) = \frac{2\pi}{\lambda} W(x)$. In our case, the index of refraction of the photoresist at 880 nm wavelength is $n$=1.63. Our phase screen (Fig. 3(a)) exhibited 3 $\mu$m peak-to-valley topography variations, corresponding to a measured phase $\sigma_\varphi$ of 4.67 radians rms. The characteristic length of the phase variations is taken to be $l_\varphi$ = 200 $\mu$m, given here by the periodicity of our aberration pattern. The correspondence between the phase screen topography and the CDM topography after AO correction is apparent (Figs. 3(a) and 3(b)), as expected since the phase screen and CDM are conjugate to one another. In contrast, the topography of the PDM after AO correction (represented in wavefront units) bears no resemblance to the phase screen topography (Fig. 3(c)), also as expected. Projection of phase-screen on pupil plane and phase wrapping in PDM produced a complex correction phase pattern as shown in Fig. 3(c). Both DMs compensate the system aberration (if present), however, such correction doesn't reduce the FOV of corrected image. The most common system aberration in such optical system is spherical aberration. The phase maps in Figs. 3(b) and 3(c) show that system correction (if present) must be comparatively smaller than the sample aberration correction. The order of aberration that conjugate AO can fix is determined by the resolution of CDM and magnification $M$ between the sample and CDM. In our present setup, aberration having characteristic length of phase variation larger than 125 $\mu m$ and peak-to-valley phase variation less than 42.8 radians can be corrected.

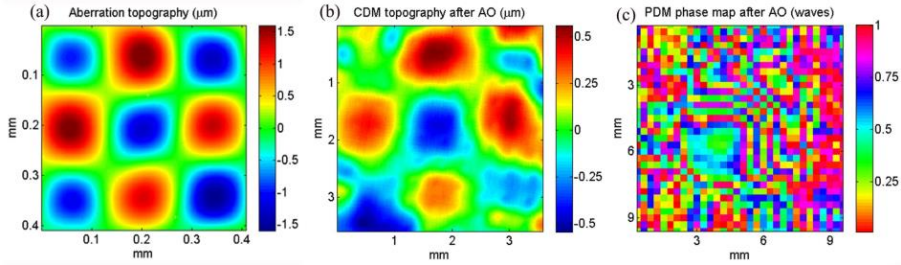

Fig. 3. (a) Topographic map of phase screen, (b) topographic map of CDM surface after AO correction, and (c) phase map of PDM after AO correction in wavefront units. Note: there is about a 7x magnification difference between the aberration plane and the conjugate CDM plane.

A benefit of including a translatable carriage in our setup is that it allowed us to conjugate the CDM to arbitrary planes in the vicinity of the sample. As such, we were able to study the axial range of our conjugate AO correction. Specifically, we first optimized our AO correction when the CDM was properly conjugated to the phase screen (i.e. 300 $\mu$m from the focal plane). Once optimized, we held the resultant CDM correction pattern fixed. We then translated the CDM to gauge the axial range of this correction, using the averaged square root of the image intensity as a quality metric. The results are shown in Fig. 4, where the physical displacement of the CDM has been translated to an effective displacement about the aberration plane (i.e. the physical displacement has been divided by $M^2$).

The axial range results in Fig. 4 may be understood from simple arguments. To begin, let us consider a perfectly conjugated CDM, and denote the optimal aberration correction it imparts as $\varphi(x)$. Before considering axial displacements, let us consider a lateral displacement of this optimal correction, denoted as $\varphi(x+\delta x)$. The resultant rms error associated with the aberration correction is then $\sqrt{\overline{|\varphi(x+\delta x)-\varphi(x)|^2}}$, averaged over all positions *x*. The aberration correction fails when this rms error reaches a certain threshold, say 1 radian. We find then that the maximum tolerance of the aberration correction to lateral displacements is roughly defined by $|\nabla\varphi(x)|\delta x_{max} \approx 1$, where $\nabla\varphi(x)$ is a characteristic slope of the aberration phase variations, leading to $\delta x_{max} \approx l_\varphi/\sigma_\varphi$. As pointed out in [31] by Mertz et al, this same maximum tolerance also corresponds to the FOV radius of pupil AO correction (a more accurate calculation√ for Gaussian phase variations yields $\delta x_{max} \approx l_\varphi/2\sigma_\varphi$ [26], which in our case corresponds to about 30 $\mu$m, in rough agreement with the results shown in Figs. 2(c) and 2(f) (the full FOV diameter is twice this value)).

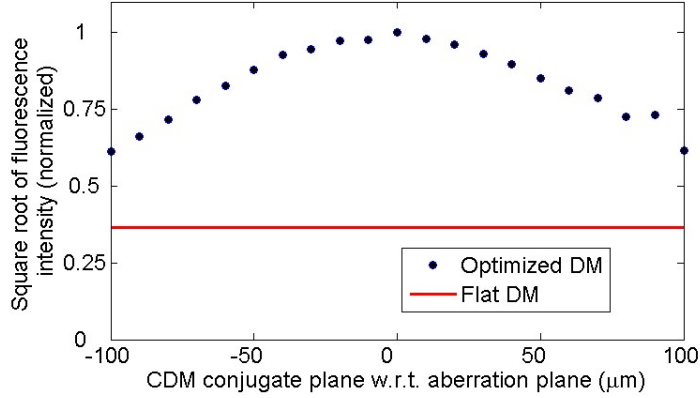

Fig. 4. Normalized averaged square root of fluorescent intensity of images versus axial translation of the CDM conjugated plane. The straight red line indicates the average square root of fluorescence intensity without conjugate AO correction.

We turn finally to a consideration of axial displacements of the aberration correction $\varphi(x)$. The light propagating through the correction plane does so with an angular diversity characterized by $\frac{2}{3}$NA, where NA is the numerical aperture of the illumination optics, and the factor of $\frac{2}{3}$ is included to account for angular averaging in a cylindrically symmetric geometry. Because of this angular diversity, axial displacements $\delta z$ of the correction plane, upon light propagation (forward or backward) to the aberration plane, cause the correction $\varphi(x)$ to exhibit translational diversity characterized by $\delta x \approx \frac{2}{3}\text{NA}\delta z$. We thus find that the maximum tolerance of the aberration correction to axial displacements is very roughly given by $\delta z_{max} \approx \frac{3}{2}\delta x_{max}/\text{NA}$. In our case, the illumination NA was close to, though a bit less than, the NA of the microscope objective because of beam underfilling, obtaining $\delta z_{max} \approx 60$ $\mu$m, in rough agreement with the HWHM of the plot shown in Fig. 4 (the full axial translation range is twice this value).

## 4. Discussion

We have demonstrated the feasibility of conjugate AO in a two-photon microscope configuration. As demonstrated previously in widefield microscopy, the compensated FOV achieved with conjugate AO in two-photon scanning microscopy is significantly larger than the corresponding compensated FOV achieved with pupil AO. The lateral range of the AO correction depends only on properties of the aberration itself (namely on the characteristic slope of the aberrating features), whereas the axial range also depends on the microscope NA, and is greater than the lateral range by a factor of about $NA^{-1}$. In our case, this range extended to more than a hundred microns axially, promising practical benefits in deep-tissue biological imaging despite the presence of interface aberrations.

Implementation of conjugate AO in a scanning microscope is relatively straightforward, but practical limitations constrain the technique. First, the FOV advantage in conjugate AO comes with an inherent compromise in the spatial resolution of AO compensation in comparison to pupil AO with the same DM. In pupil AO, the position of the excitation beam is fixed on the DM aperture independently of beam scanning, whereas in conjugate AO, it translates within the DM aperture. Pupil AO can thus employ the entire DM aperture to compensate pupil aberrations, while conjugate AO employs DM sub-apertures corresponding to different scan positions. When the DM is properly conjugated to the aberration plane, the collection of these scanned subapertures fills the entire DM aperture, but for any particular scan position the subaperture comprises fewer spatial degrees of freedom than the DM has available in total. This trade-off between compensation spatial resolution and corrected FOV must be considered in optical system design to optimize AO performance based on the expected character of the sample aberrations and the requirements of the imaging task.

A second challenge comes from the AO feedback mechanism itself. Here, we employed stochastic perturbation of the DM and a gradient descent optimization technique based on image quality (here characterized by total intensity). This approach suffers from two major drawbacks. The first is that it is slow, requiring hundreds of iterations to compensate a given aberration and making it difficult to implement in real time. The second is that convergence is not guaranteed, and even when the AO loop does converge, there is no guarantee that the solution is globally optimal. The success of AO optimization based on image intensity metrics in two-photon microscopy is strongly dependent on the properties of the object being imaged, including sparsity of fluorescent emitters and their susceptibility to photobleaching. Moreover, image-based optimization metrics can fail in deep-tissue two-photon microscopy because of the relatively low levels of signal to background.

Finally, in our demonstration of conjugate AO we limited ourselves to a single-layer sample and a single layer aberration. While such a geometry can be encountered in practice, it is by no means general [34]. For example, let us consider the possibility that the sample is axially extended. This does not present a fundamental issue for a two-photon microscope since the fluorescence excitation is inherently limited to a single layer, namely the focal plane. Nevertheless, to image a volumetric sample one must acquire an image stack, meaning that the axial separation between the focal plane and the aberration plane must vary during the course of acquisition. Accordingly, the DM must be translated to remain conjugate with the aberration plane. Our motorized translation stage shown in Fig. 1 was designed to do just this, but only to a limit. In general, for a change in axial separation between the object and aberration of $\Delta z$, the DM must be translated axially by a distance $M^2 \Delta z$ to remain conjugate with the interface aberration plane. This distance can rapidly become impracticable and impose a constraint on the achievable axial range of volumetric imaging, especially in systems with high magnification. For example, in our proof-of-principle apparatus with magnification ~6.4×, we were limited to an axial scan range corresponding to $\Delta z \approx 300$ $\mu$m. We note that in the case where the separation of the object and the aberration remains fixed, the DM position for conjugate AO also remains fixed and much of the complexity of the optical layout shown in Fig. 1 can be eliminated.

A more fundamental limitation comes from situations where the aberrations themselves are not confined to a single layer but rather distributed throughout the sample volume. While comparable problems in astronomical imaging have been overcome successfully with multiconjugate AO [23–25], the question remains to what degree singly-conjugate AO can achieve similar success. Numerical simulations have shown that benefits of conjugate AO persist even when only a single DM is employed [27, 29, 30]. Our experimental results suggest this is indeed the case. Specifically, they show that conjugate AO correction is relatively long range in the axial direction, particularly in the case of modest to low NA. Such long range correction implies that a single DM correction can serve to compensate, at least partially, a commensurate axial range of volumetric aberrations. While it remains to be seen how well the approach demonstrated here will work in actual biological imaging applications of interest, preliminary indications appear encouraging.

**Acknowledgments**

Grant support for this project was provided by National Science Foundation Industry/University Cooperative Research Center for Biophotonic Sensors and Systems (IIP-1068070). Loaned equipment support was provided by Thorlabs Corporation and by Boston Micromachines Corporation. Professor Bifano acknowledges a financial interest in Boston Micromachines Corporation, which manufactures and sells the deformable mirrors used in this work.